\documentclass[12pt]{article}

\usepackage{epsf}
\usepackage[dvips]{graphicx}

\begin{document}
\begin{center}
{\bfseries Multiplicity and angular distribution of particles
emitted in relativistic nuclear-nuclear interactions }

\vskip 5mm

M. K. Suleymanov${1,2\dag}$, E. U. Khan$^{1}$, A Kravchakova$^{3}$ ,
Mahnaz Q. Haseeb$^{1}$,S. M. Saleem$^{1}$, Y. H.
Huseynaliyev$^{1,4}$, S Vokal$^{2,3}$, A.S. Vodopianov$^{2}$, O.B.
Abdinov$^{5}$

\vskip 5mm

{\small (1) {\it Department of Physics COMSATS Institute of
Information Technology, Islamabad, Pakistan }
\\
(2) {\it Veksler and Baldin Laboratory of High-Energies, JINR,
Dubna, Russia}
\\
(3) {\it  University of P. J. Shafarik, Koshice, Slovak Republic}
\\
(4) {\it Department of Semiconductor Physics, Sumgayit State
University, Azerbaijan}
\\
(5) {\it Physics Institute of AS, Baku, Azerbaijan}
\\
$\dag$ {\it E-mail: mais@jinr.ru}}
\end{center}

\vskip 5mm

\begin{center}
\begin{minipage}{150mm}
\centerline{\bf Abstract} We discuss the experimental results on the
behavior of the average multiplicities and angular distributions of
slow particles emitted in hadron-nuclear and nuclear-nuclear
interactions at relativistic energies as a function of the
centrality of collisions. It is observed that by increasing the mass
of the projectiles the angular distributions of slow particles
change and the structure which was demonstrated in the case of
$\pi$-mesons, protons and light nuclear projectiles, almost
disappears. During the interaction of the heavier projectile with
nuclear target, the number of secondary interactions as well as
number of nucleon-nucleon elastic scattering and re-scattering
events increases. We suggest to restore this information using the
heavy ion generators taking into account the multiplicity
distributions. Because our investigations show that the formation of
the percolation cluster sufficiently influences the behaviour of the
average multiplicity of the slow particles emitted in these
interactions.
\end{minipage}
\end{center}

\vskip 10mm

\section{Introduction}Observation of the effects connected with formation
and decay of the percolation clusters in heavy ion collisions at
ultrarelativistic energies could be the first step for getting the
information about deconfinement. Ref. ~\cite{satz1} discusses in
detail that deconfinement is expected when the density of quarks and
gluons becomes so high that it is no longer possible to separate
color-neutral hadrons, since these would strongly overlap. Instead
clusters much larger than hadrons are formed, within which color is
not confined; deconfinement is thus related to cluster formation.
Such signals may be obtained by studying the behavior of the
characteristics of secondary particles as a function of the
centrality of collisions, mass and energy of the projectiles and
targets. The formation and decay of the intermediate baryons may
influence the characteristics of secondary particles, especially
those of slow particles. Ref.~\cite{agnese1} discusses that the
angular distributions of fragments could have special structure as a
result of the formation and decay of some intermediate states.

What we have got from experiments?  There are only a few
experimental results~\cite{anoshin1}-\cite{andreeva1}  that can be
considered as a possible confirmation of it. In Ref.
~\cite{anoshin1} it was obtained that the angular distributions
of protons emitted in $\pi^{-12}C$-interaction (at 40 GeV/c) with
total disintegration of nuclei have some structure, pick at angles
close to $60^0$ (see Fig.1). This result was confirmed by the data
which were obtained at investigation of the angular distributions of
protons emitted in $\pi^{-12}C$-interaction (at 5 GeV/c) with total
disintegration of nuclei~\cite{abdinov1}(see Fig.2). There were
several speculations interpret the result but it could not been understood
completely.

Earliest on Ref.~\cite{hecman1} was presented the angular
distribution for slow protons emitted in the central $He+Em$- ( at
2.1 A GeV), $O+Em$ - (2.1 A GeV) and $Ar+Em$ - (1.8 A GeV)
collisions (see Fig.3). Some wide structure was observed in these
distributions.

Later  in Ref.~\cite{andreeva1} reported observation of some wide
structure in the angular distributions of the b-particles emitted in
the $Ne+Em$ reactions at 4.1 A GeV (see Fig.4-5). It was commented
that the structure becomes cleaner in central collisions.

The major interpretation from these experiments is that the angular
distributions of slow particles emitted in $\pi$-meson, proton and
light nuclei interactions with nuclear targets indicate a structure.
The reason of this structure could be the formation and decay of an
intermediate baryon system, e.g. percolation clusters
(see~\cite{volant1}).

Now what we have got for angular distributions of slow particles
emitted in intermediate and heavy ion collisions? Up to now those
distributions could not show any structure. Why? May be it is
because intermediate baryon systems have a small cross section
(statistical reason)  or the information on these formations is lost
for reasons, such as the secondary multiparticle interaction,
scattering and rescatterings (dynamical reason). We can exclude the
statistical reason because the structure for slow particle's angular
distributions has been observed with lesser statistical data. What
about dynamical reason?

To get an answer to this question we study in detail the behavior of
average multiplicity and angular distribution of slow particles
emitted at intermediate mass and heavy nuclear interaction with
emulsion target at relativistic energies as a function of
centrality.

\section{ Experiment}To analyze the behaviour of average multiplicity and
angular distribution of the $b$ -particles (mostly protons with $p =
0.2 GeV/c$ and multiple charged target fragments having a range,
3mm) as a function of centrality~\cite{suleyman1} the experimental
data on $Kr+Em$ -reaction at 0.95 A GeV~\cite{krasnov1} and $Au+Em$
- reaction at 10.7 A GeV~\cite{adamovich1} has been used. This data
was obtained using nuclear beams of the SIS GSI and AGS BNL by the
EMU01 Collaboration. In all 842 $Kr+Em$ events and 1185 $Au+Em$ ones
have been considered.

Before discussing the experimental results we would like to touch
upon one more question which is very important for the centrality
experiments. It is clear that the centrality of collisions can not
be defined directly in the experiment. In different   experiments
the values of centrality is detected as a number of identified
protons , projectiles'  and targets' fragments, slow particles, all
particles, as the  energy flow of the particles with emission angles
$\theta = 0^0$   or with $\theta=90^0$ . Apparently, it is not
simple to compare quantitatively the results on centrality
dependencies obtained in different papers and on the other hand the
definition of centrality could significantly influence the final
results.  So we believe it is necessary to understand what
centrality is.  Usually for a chosen variable to fix centrality it
is supposed that its values have to increase linearly with a number
of colliding nucleons or baryon density of the nuclear matter. The
simplest mechanism that could give this dependence is the cascade
approach. So, we have used one of the versions of the
cascade-evaporation model CEM~\cite{cem} to choose the variable to
fix centrality for studying the centrality dependence of the event
characteristics.

\subsection{Average multiplicities}   Fig. 6a-c      shows the yields of
$g - , h$ - and $F$-fragments  in the  $Kr + Em$ reactions at 0.95
GeV/nucl. The results coming from the CEM  are also included. We can
see that : The g-fragments experimental multiplicity distribution
($N_g$) for the $Kr + Em$ reactions is well described by the model
(Fig. 6a). We recall that  in the framework  of the CEM
$g$-fragments are considered as the results of cascading collisions
in the target spectator and participant. So $N_g$   could be used to
fix the centrality of collisions.  The target $h$-fragments
multiplicity ($N_h$) distribution shape for the $Kr + Em$ reactions
(Fig. 6b) cannot be described by  the CEM in  the region $15 < N_h <
32$ . If we remember that  $N_h$ is the number of the final state
target fragments in the event which is the sum of the target black
fragments ($N_b$) and $N_g$, we would say that the CEM is not able
to describe the multiplicity distributions of $b$-particles which
are the slowest target fragments and thus there is a need to get
much more information on the state of the nuclear target.

Fig 7d-f shows the yields of  $g - , h$ - and $F$-fragments in the
$Au + Em$ (at 10.6 GeV/nucl) reactions. The results coming from the
CEM are included as well. The experimental distribution of
$g$-particles ( Fig. 7d)  can not be described by  the model in full
region of the $N_g$ definition. We can separate some regions with
different from of the experimental and model distributions. In the
region of $N_g < 5$ the model can describe the experimental
distribution. In the region of  $N_g > 15$ the experimental values
of $N_i$ decrease with $N_g$ while the values coming from the model
are constant  in the region $15 <  N_g <  40$. The  model could  not
describe the distribution of $h$-particles in a full region of the
$N_h$-definition either, that is seen from Fig. 7e. The model could
only describe the experimental distribution of $N_h$ in the region
of $22< N_h < 32$. Thus one can say that for the reaction  under
consideration the  $N_h$  as well as  the $N_g$ could not be used to
fix the  centrality of collisions. We believe that the result could
also be understood qualitatively in the framework of the
above-mentioned physical picture (for high energy interactions).

The distribution of projectile fragments with $Z\ge 1$ produced in
$Au + Em$ collisions is  in good  agreement with the result  coming
from the CEM (Fig.7f). So, we can see that the projectile fragments
are produced by the mechanism similar to the cascade-evaporation one
and $N_F$ may be used to fix the centrality for these reactions.

Fig. 8a-c presents the average values of multiplicity $<n_s>$ for
$s$ - particles produced in $Kr + Em$ and  $Au + Em$  reactions and
the average values of pseudorapidity for s - particles produced in
$Au + Em$ reactions . We can say that there are two regions in the
behavior of the  values  of  $<n_s>$   as a function of  $N_g$  for
the $Kr+Em$ reaction ( Fig. 8a).

In the region of : $N_g  < 40$ the values of  $<n_s>$ increase
linearly with $N_g$ , here  the  CEM also gives the linear
dependence but with the slope  less than the experimental one; $N_g
> 40$ the CEM gives the values for $<n_s>$   greater than  the
experimental observed ones, the last saturates in this region. This
effect could not be described by the CEM . It has been however,
previously observed in emulsion experiments. It is clear that there
should be some  effects which could stop ( or sufficiently moderate)
the increase of  $n_s$. The effect of the big percolation cluster
formation could be one of those effects. The moderation of the
values of $<n_s>$ as a function of $N_F$ is also observed for the
$Au+Em$ reaction at 10.6 GeV/nucl. (Fig. 8b) near the point of the
$N_F\ge  40-50$ these should be a point of the regime change which
is absent for the distribution coming from the CEM. Thus our
investigations show that ~\cite{abdinov2}:

       - the centrality of collision could be defined  as a number of  the  target g-particles
       (grey particles are those whose ionization, the number of grains per unit length, correspond
       to protons with momentum $0.2\le p \le 1.0~ GeV/c$) in $Kr + Em$ reactions at energies 0.95 A GeV/nucl
       and as a number of projectile F-fragments (with $Z\ge 1$)  in $Au + Em$ reactions at 10.6 A GeV/nucl;

      - the formation of the percolation cluster sufficiently influences the characteristics of nuclear fragments;

      - there are points of the regime changes in the behavior of some   characteristics of s-particles as a function
      of centrality which could be qualitatively understood as a result of big percolation cluster formation.

\subsection{Angular distributions} The angular distributions of slow particles
were presented by the EMU01 Collaboration (for example
see~\cite{adamovich2}-\cite{adamovich3}. However the angular
distributions of the slow fragments are considered by us separately
for:

1) all events and the events with a number of $N_h = 8$ ($N_h$ is a
number of $h$-particles). This criterion has been used to separate
the heavy nuclei interaction. The $h$ -particles are sum of $g$  and
$b$ -particles;

2)  peripheral and central collisions.

The experimental data has been compared with data generated by the
CEM.

Fig 9 a-d shows the angular distributions of the $b$-particles
emitted in the $Kr + Em$ - reactions at 0.95 A GeV. Herein and
further all distributions were normalized on a total square under
the curves. The results are shown separately for all events (Fig.9a)
and the events with $N_h = 8$ (Fig.9b). The results from the CEM are
also shown in these figures. There is not seen any structure in
these distributions such as those discussed in
~\cite{anoshin1}-\cite{andreeva1}.

In Fig. 10 a-d the angular distributions of the $b$- particles
emitted in the $Au + Em$ - reactions at 10.7 A GeV are shown. The
results are demonstrated separately for all events (Fig.10a) and the
events with $N_h = 8$ (Fig.10b). These figures show some structure
in the angular distributions for b-particles that is cleaner for all
events for which the rates of the peripheral collisions are
stronger. The possible reasons for the appearance of this structure
are the multinucleon interactions, scattering and rescatterings
effects.

In Fig. 11 a-b the angular distributions of the $b$- particles
emitted in the $Kr + Em$ - reactions at 0.95 A GeV are demonstrated.
The results for the central collisions (Fig. 11a) and the peripheral
ones (Fig. 11b) are shown separately. To select the central
collisions, we used the criteria $N_g = 20$ suggested by Abdinov et
al~\cite{abdinov2}. One can see that the angular distributions are
completely described by the CEM.

In the Fig 12 a-b the angular distributions of the b-particles
emitted in the $Au + Em$ - reactions at 10.7 A GeV are demonstrated.
The results for the central collisions (Fig 12a) and the peripheral
ones (Fig 12b) are shown separately. To select the central
collisions, we used the criteria $N_F = 20$ though Abdinov et
al~\cite{abdinov2} showed that the central Au+Em-events (at 10.7 A
GeV) should be selected using the criteria $N_F = 40$. We followed a
different criterion as the number of events is very small with $N_F
= 40$. It is seen that in peripheral collisions the structure
becomes cleaner (the structure diminishes and almost disappears in
central collisions). Therefore as in Fig 10a-b one can say that the
reason of the appearance of this structure is the multinucleons
interactions, scattering and rescattering effects.

\section{Results and discussion}The angular distributions of
$b$-particles emitted in $Au+Em$-reactions as well as in the
$Kr+Em$-reactions do not contain any special structure except the
one that is described using the usual mechanisms of the interaction
(e.g. cascade evaporation). If we compare the results of the
reactions mentioned above with those discussed
in~\cite{anoshin1}-\cite{andreeva1} one can see that by increasing
the mass of the projectiles the angular distributions of slow
particles change and the structure which was demonstrated in the
case of $\pi$-mesons, protons and light nuclear projectiles,
diminishes. This could be explained easily as follows: during the
interaction of the heavier projectile with nuclear target, the
number of secondary interactions as well as number of
nucleon-nucleon elastic scattering and re-scattering events
increases. These effects could lead to the disappearance of the
information of any intermediate formations as well as the clusters.

    But how could we restore this information? As already discussed,
    observation of the effects connected with formation and decay of the
    percolation clusters in heavy ion collisions at ultrarelativistic
    energies could be the first step for getting the information about
    deconfinement.  We expected that such signals may be obtained by
    studying the behaviour of the characteristics of secondary particles
    as a function of the centrality of collisions, mass and energy of the
    projectiles and targets. But these could not be observed.
            Therefore it is necessary to look for the ways for restoring
            the signal. One way could be using the generators of heavy ion
            events taking into account the results on the centrality dependences
            of other characteristics of secondary particles e.g., the average multiplicity
            of b-particles~\cite{abdinov2}
            which could be more sensitive to baryon system formation and decay.
            In Ref.~\cite{abdinov2} it was
            concluded that the formation of the percolation cluster could sufficiently
            influence  the behavior of the  average multiplicity of slow particles as a
            function of centrality. Another way could be using the Fourier transformation
            and Maximum Entropy technique ~\cite{belashev123} to separate the
            spectrum of the signal. It is the subject of our investigations that would follow
            in near future.

%% To insert figure (with the help of epsf.sty)
\begin{figure}[h]
 \centerline{
\includegraphics[width=120mm,height=80mm]{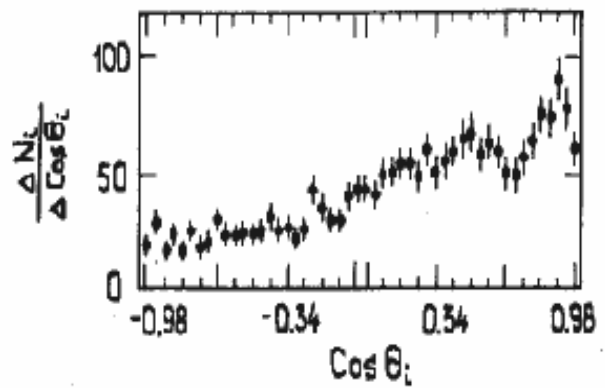}}
 \caption{The angular distributions of protons emitted in $\pi^{-12}C$-
interaction (at 40 GeV/c) with total disintegration of nuclei.}
\end{figure}

\begin{figure}[h]
 \centerline{
\includegraphics[width=60mm,height=60mm]{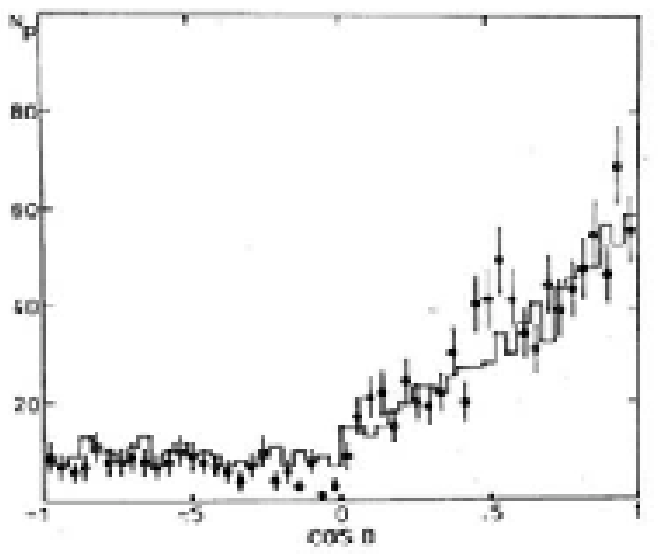}}
 \caption{The angular distributions of protons emitted in $\pi^{-12}C$-
                interaction (at 5 GeV/c) with total disintegration of nuclei.}
\end{figure}

\begin{figure}[h]
 \centerline{
\includegraphics[width=50mm,height=100mm]{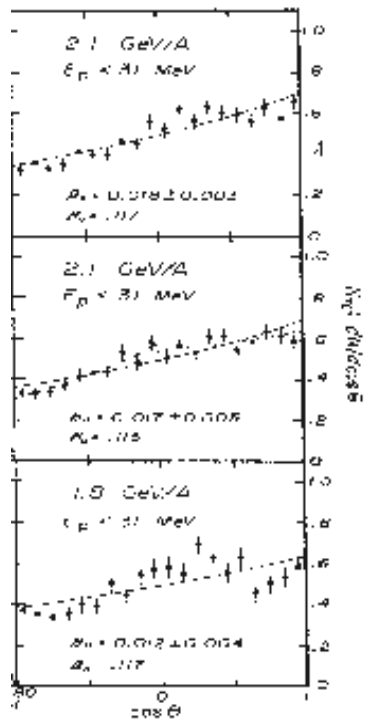}}
 \caption{The angular distributions for slow protons emitted in the central
                    He+Em-( at 2.1 A GeV), O+Em - (2.1 A GeV) and Ar+Em - (1.8 A GeV)
                    collisions.}
\end{figure}
\begin{figure}[h]
 \centerline{
\includegraphics[width=60mm,height=50mm]{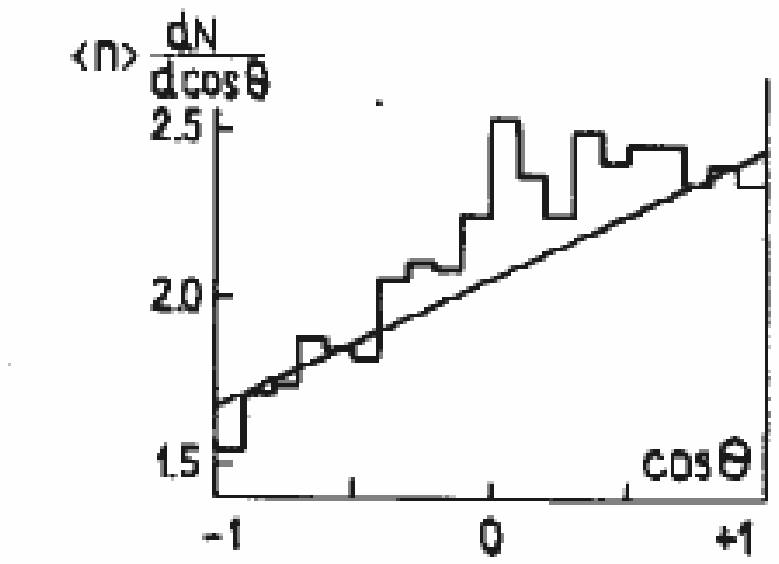}}
 \caption{The angular distributions of the b-particles emitted in  the
                    Ne+Em reactions at 4.1 A GeV .}
\end{figure}

\begin{figure}[h]
 \centerline{
\includegraphics[width=100mm,height=80mm]{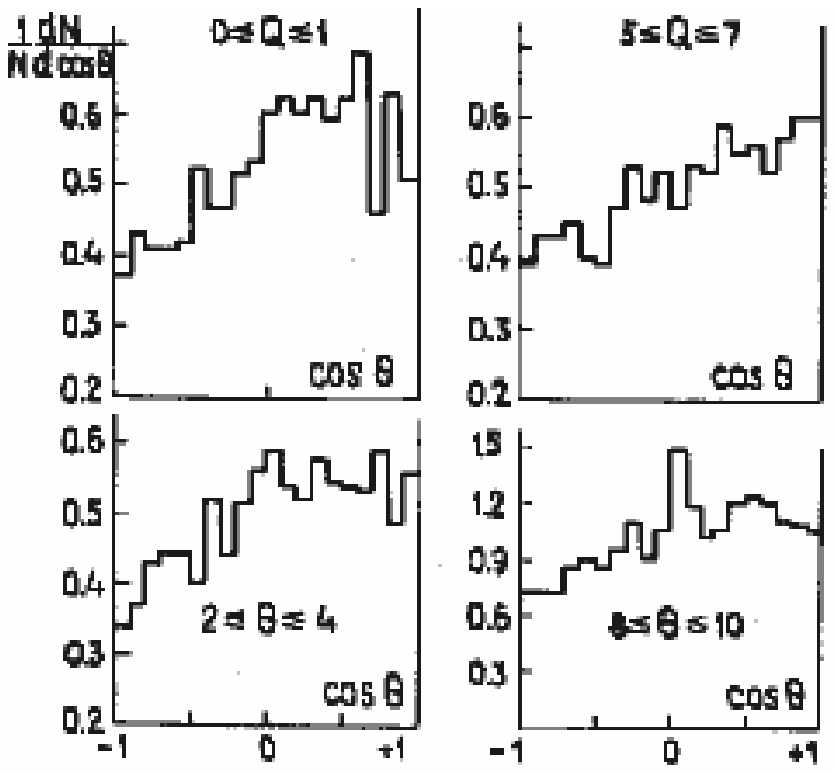}}
 \caption{The angular distributions of the b-particles emitted in  the
                    Ne+Em reactions (with different centralities Q) at 4.1 A GeV .}
\end{figure}
\begin{figure}[h]
 \centerline{
\includegraphics[width=100mm,height=110mm]{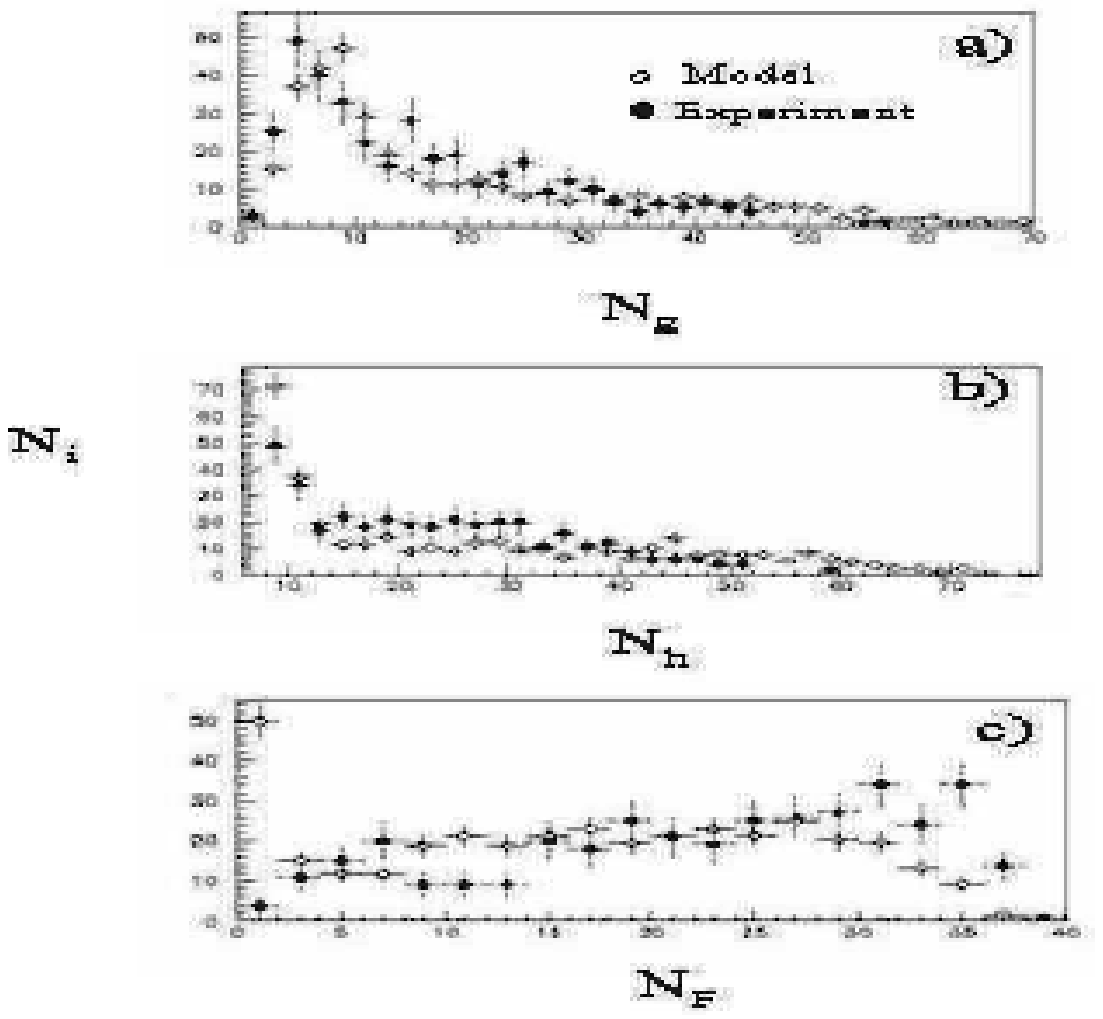}}
 \caption{The yields of
$g - , h$ - and $F$-fragments  in the  $Kr + Em$ reactions at 0.95
GeV/nucl. The results coming from the CEM  are also drawn. }
\end{figure}

\begin{figure}[h]
 \centerline{
\includegraphics[width=100mm,height=110mm]{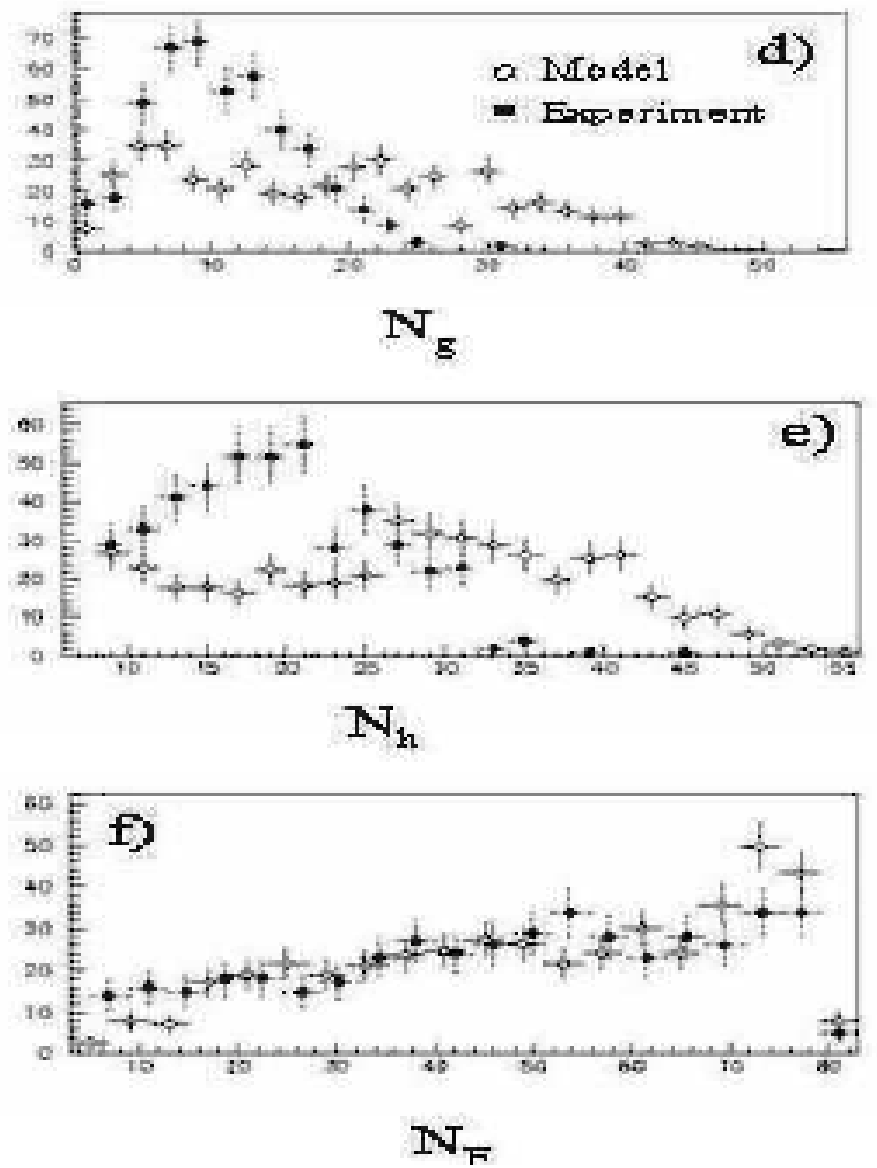}}
 \caption{The yields of
$g - , h$ - and $F$-fragments  in the  $Au + Em$ reactions at 10.7
GeV/nucl. The results coming from the CEM  are also drawn. }
\end{figure}

\begin{figure}[h]
 \centerline{
\includegraphics[width=150mm,height=60mm]{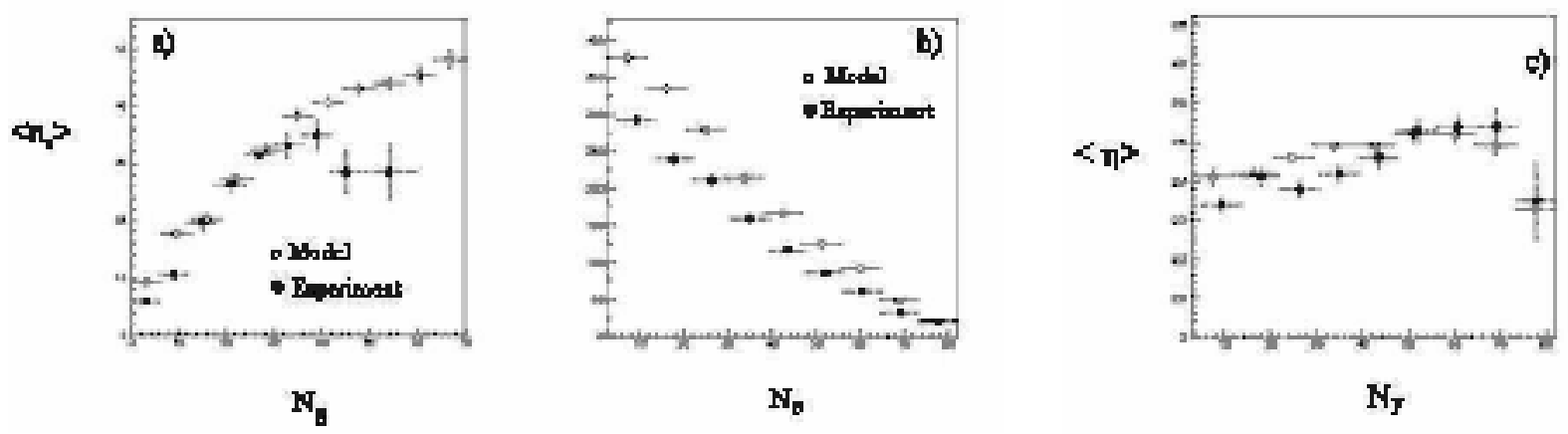}}
 \caption{}
\end{figure}

\begin{figure}[h]
 \centerline{
\includegraphics[width=110mm,height=40mm]{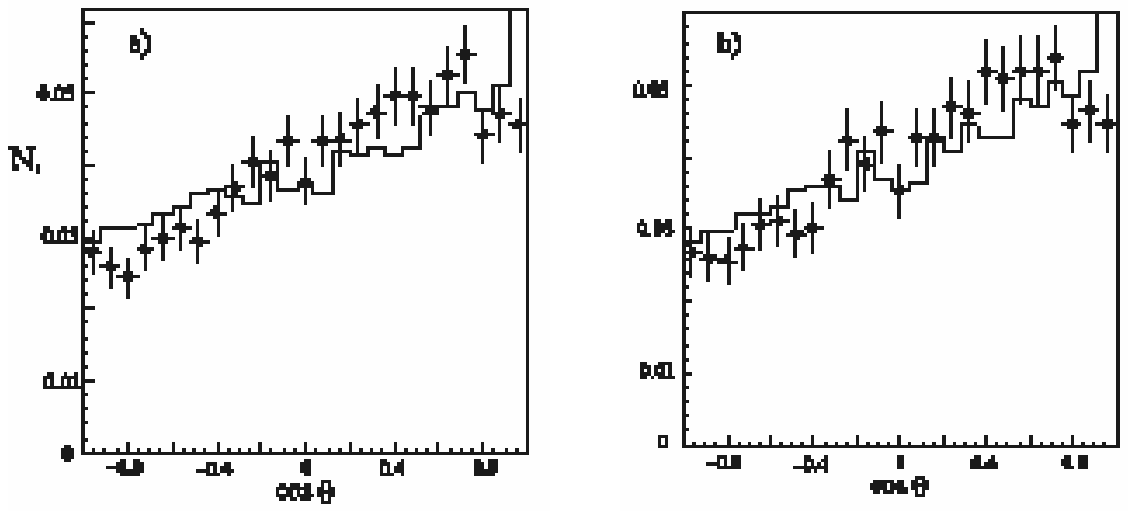}}
 \caption{Angular distributions of the b- particles emitted in Kr+Em-reactions at 0.95 A GeV energy  for all events - a)
 ;  for  the events with Nh = 8 - b). The histograms are the results coming from the CEM.}
\end{figure}

\begin{figure}[h]
 \centerline{
\includegraphics[width=110mm,height=40mm]{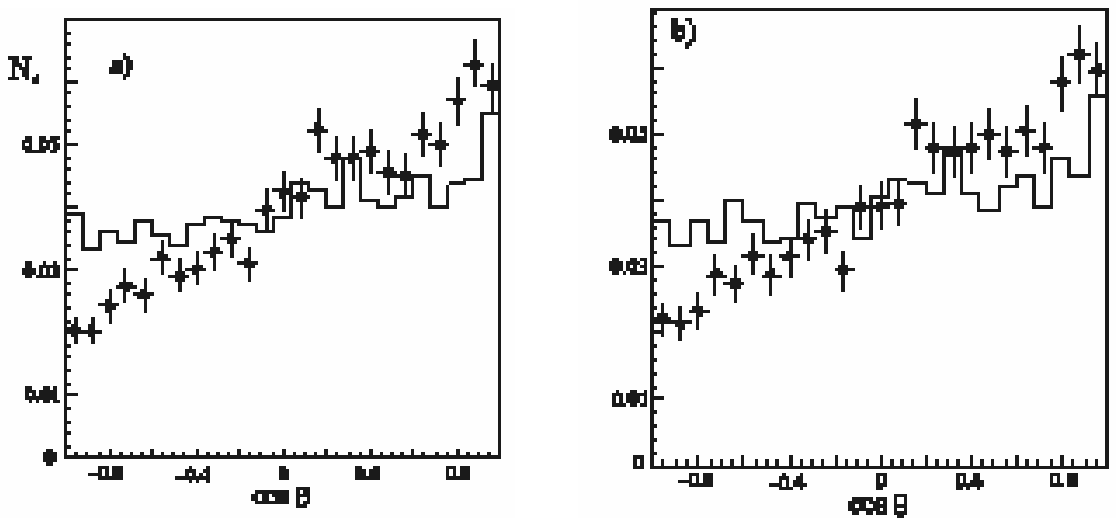}}
 \caption{Angular distributions of the b-particles emitted in Au+Em-reactions at 10.7 A energy for all events - a) ;
 for  the events with Nh = 8 - b). The histograms are the results coming from the CEM.}
\end{figure}

\begin{figure}[h]
 \centerline{
\includegraphics[width=110mm,height=40mm]{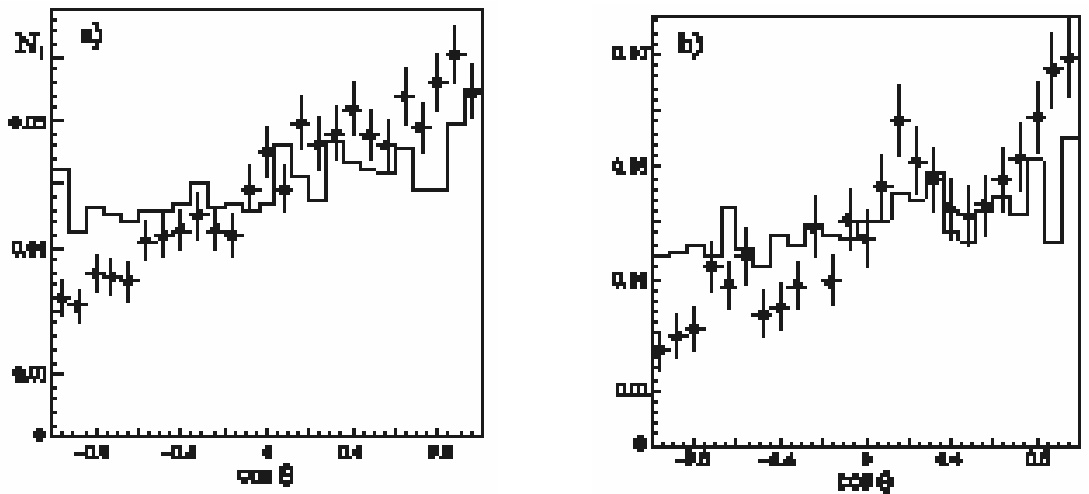}}
 \caption{Angular distributions of the b- particles emitted in Kr+Em-reactions at 0.95 A GeV energy
for  the central collisions  - a) ;  for  the  peripheral collisions
- b). The histograms are the results coming
                from the CEM.}
\end{figure}

\begin{figure}[h]
 \centerline{
\includegraphics[width=110mm,height=40mm]{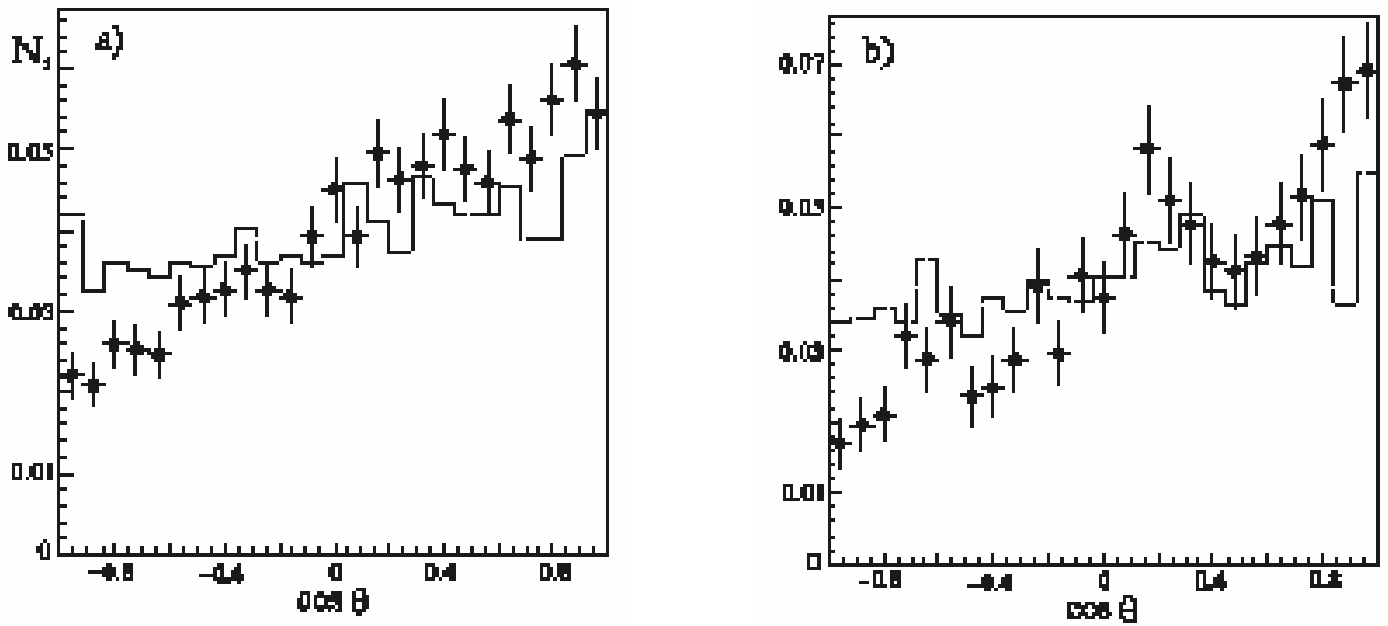}}
 \caption{Angular distributions of the b-particles emitted in Au+Em-reactions at 10.7 A GeV energy
                for  the central collisions  - a) ;  for  the  peripheral collisions - b). The histograms are the results coming
                from the CEM.}
\end{figure}


\begin{thebibliography}{99}

\bibitem{satz1}H.Satz. Preprint 2000 hep-ph/0007069].

\bibitem{agnese1} A. Agnese , M. La.Camera,A.Wataghin. Nuovo Cim. A 59 71, 1969.

\bibitem{anoshin1} Anoshin A I et al 1981 Rus. J. of Nucl.Phys. 33
164.

\bibitem{abdinov1} Abdinov O B et al 1980 Preprint JINR 1-80-859.

\bibitem{hecman1} Hecman H H et al 1978 Phys.Rev. C 17.

\bibitem{andreeva1} Andreeva N P et al 1987 Rus. J. of Nucl.Phys 45
123.

\bibitem{volant1} Volant C et al 2004 Nucl. Phys A 734 545

\bibitem{suleyman1} Suleymanov M K et al
2004 Nucl.Phys. A 734S 104

\bibitem{krasnov1} Krasnov S A et al 1996 Czech. J. Phys. 46 531

\bibitem{adamovich1} Adamovich M I et al 1995 Phys. Lett. B 352 472

\bibitem{cem} Musulmanbekov G J 1992
(11th EMU01 Collaboration Meeting) 288; Musulmanbekov G J 1994 (11th
Int. Symp. on High Energy Spin Phys.); Musulmanbekov G J 1995 (AIP
Conf. Proc.) 343 428

\bibitem{abdinov2} O.B. Abdinov et al. Journal Bulletin of the Russian
Academy of Sciences. Physics , 2006, v.70,N5, pp.656-660; e-Print
Archive: hep-ex/0503032

\bibitem{adamovich2} Adamovich M I et al 1995
Nucl.Phys. A 590 597

\bibitem{adamovich3} Adamovich M I et al 1997 Z.Phys. A 358 337

\bibitem{belashev123}Belashev B.Z., Suleymanov M.K.  .Particles and
            Nuclei, Letters. 2002. No. 3[112]; No.   4[113], ; No. 6[115].

\end{thebibliography}
\end{document}